\documentclass{article}
\usepackage{graphics,graphicx,array,hhline,color,float,mdwlist,amsmath,amssymb,cite,multirow,xcolor,hyperref,tabularx,placeins,xspace}
\usepackage[left=2cm,right=2cm,top=2cm,bottom=2cm,head=1cm,foot=1cm]{geometry}
\graphicspath{{./figs/}}
\usepackage{t1enc}
\usepackage[latin2]{inputenc}

\newcommand{\sqsntwo}{\mbox{$\sqrt{s_{_{NN}}}=200$~GeV}\xspace}
\newcommand{\auau}{\mbox{Au$+$Au}\xspace}
\newcommand{\mT}{m_{T}}

\begin{document}
\title{Femtoscopy via L\'evy sources with PHENIX at RHIC}
\author{M\'at\'e Csan\'ad\footnote{Presented at the XXXVII International Symposium
on Physics in Collision} \; for the PHENIX Collaboration\\
E\"otv\"os Lor\'and University, H-1117 Budapest, P\'azm\'any P. s. 1/A, Hungary}
\maketitle

\begin{abstract}
Charged pion two-particle correlation functions were measured in $0-30\%$ centrality $\sqrt{s_{_NN}}=200$ GeV Au+Au collisions
with the PHENIX experiment at RHIC. The measured correlation functions can be statistically well described based on the
assumption of Lévy-shaped source distributions. In this proceedings paper we present the Lévy parameters of the
measured correlation functions: correlation strength parameter $\lambda$, Lévy index $\alpha$ and Lévy scale
parameter $R$ as a function of pair transverse mass $\mT$, in 31 bins from 228 to 871 MeV, separately for
positive and negative pion pairs. We discuss the physical interpretation of the $\mT$ dependence of the parameters.
\end{abstract}

\section{Introduction}
In ultra-relativistic collisions of heavy ions, strongly coupled Quark Gluon Plasma (sQGP)
is formed~\cite{Adcox:2004mh,Adams:2005dq,Arsene:2004fa,Back:2004je} for a very short amount of time,
and after a quark-hadron freeze-out, hadrons are created. The measurement of Bose-Einstein correlations (i.e. femtoscopy)
can be used to gain knowledge about the space-time geometry of the particle emitting source,
as originally observed by~\cite{Goldhaber:1959mj,Goldhaber:1960sf}, and in radio and optical
astronomy by R. Hanbury Brown and R. Q. Twiss (HBT)~\cite{HanburyBrown:1956bqd}.
In an interaction-free case, the two-particle Bose-Einstein correlation functions are related to the Fourier
transform of the source function ($S(x,k)$, describing the probability density of particle creation
at the space-time point $x$ and with four-momentum $x$):
\belowdisplayskip=5pt
\abovedisplayskip=5pt
\begin{align}
C^{(0)}_2(Q,K)&\simeq 1+\left|\frac{\widetilde{S}(Q,K)}{\widetilde{S}(0,K)}\right|^2,\label{e:C0}
\end{align}
where $\widetilde{S}(q,k)=\int S(x,k)e^{iqx}d^4x$ is the Fourier-transformed of $S$,
and $Q = p_1 - p_2$ is the momentum difference, $K = (p_1+p_2)/2$ is the average
momentum, and we assumed, that $q \ll K$ holds for the investigated kinematic range.
Usually, correlation functions are measured versus $Q$, for a well-defined $K$-range,
and then properties of the correlation functions are analyzed as a function of the average $K$
of each range. In an expanding Gaussian source, then $1+\exp -(QR)^2$ correlations 
are thus measured, where the observed Gaussian radius $R$ is a homogeneity length,
depending on the average momentum $K$ or the related transverse mass $\mT$. The approximate dependence
of  $R^{-2}\propto a+b \mT$ is observed, rather universally (for various collision systems,
collision energies and particle types)~\cite{Adler:2004rq,Afanasiev:2009ii}, which 
can be interpreted as a consequence of hydrodynamical expansion~\cite{Makhlin:1987gm,Csorgo:1995bi}.
See Ref.~\cite{Adare:2017vig} (and references therein) for details.

It is important to note, that a significant fraction of pions are secondary, coming from decays.
Hence the source will have two components: a core of primordial pions, stemming from the
hydrodynamically expanding sQGP, and a halo, consisting of the decay products of long lived resonances
(such as $\eta$, $\eta'$, $K^0_S$, $\omega$): $S=S_{\rm core}+S_{\rm halo}$.
These two components have characteristically different sizes ($\lesssim 10$ fm for the core,
$>50$ fm for the halo, based on the half-lives of the above mentioned resonances).
In particular, the halo component is so narrow in momentum-space,
that it cannot be resolved experimentally. This leads to the following apparent
correlation function:
\begin{align}
\lim\limits_{q\rightarrow 0} C^{(0)}_2(Q,K) = 1 + \lambda(K),
\end{align}
where $\lambda = N_{\rm core}/(N_{\rm core}+N_{\rm halo})$ was introduced,
being related related to the fraction of primordial
pions among all (primordial plus decay) pions. One of the motivations for measuring
$\lambda$ is that it is related~\cite{Vance:1998wd} to the $\eta'$ meson yield,
expected~\cite{Kapusta:1995ww} to increase in case of chiral $U_A(1)$ symmetry
restoration in heavy-ion collisions (due to the expected in-medium mass decrease of the $\eta'$).
Note that a study~\cite{Csorgo:2009pa} reported the compatibility of existing $\lambda(\mT)$ data and predictions
based on a decreased in-medium $\eta'$ mass.

Experimental results show~\cite{Afanasiev:2007kk,Adler:2006as}, that the source function does 
not always exhibit a Gaussian shape. In an expanding hadron resonance gas, increasing mean free paths lead to a Lévy-flight,
anomalous diffusion, and hence to spatial Lévy distributions~\cite{Metzler:1999zz,Csorgo:2003uv,Csanad:2007fr}
This leads to a correlation function of
\begin{align}
C^{(0)}_{2}(Q,K)=1+\lambda(K)\cdot e^{-(QR(K))^{\alpha(K)}},
\end{align}
where $\alpha$ is the ($K$-dependent) Lévy-exponent, which is conjectured~\cite{Csorgo:2009gb} to be
identical to the critical exponent $\eta$, conjectured to take a value of
0.5 or even lower, identivally to the universality class of the 3D Ising model
(possibly with random external fields)~\cite{El-Showk:2014dwa,Rieger:1995aa,Halasz:1998qr,Stephanov:1998dy,Csorgo:2009gb}.
Since the exploration the search for the QCD critical endpoint is one of the major goals of
experimental heavy ion physics nowadays, we gain additional motivation for the measurement
and analysis of of Bose-Einstein correlation functions.

%We also have to take into account that the interaction-free case is not valid for the usual measurement of
%charged particle pairs, the electromagnetic and strong interactions distort the above simple picture.
%For identical charged pions, the Coulomb interaction is the most important, and this decreases the number of particle
%pairs at low momentum differences. This can be taken into account by utilizing the Coulomb pair wave function
%solving the Schrödinger-equation for charged particles, given for example in Ref.~\cite{Adare:2017vig}.
%With this, a so-called ``Coulomb-correction'' introduced, and the measured correlation function can be described by
%\begin{align}\label{e:KCoulomb}
%C_2^{\rm measured}(q,K) = K_2(q,K)C^{(0)}_2(q,K).
%\end{align}
%For details, see again Ref.~\cite{Adare:2017vig} and references therein.

Hence, in the following we utilize a generalization of the usual Gaussian shape of
the Bose-Einstein correlations, namely we analyze our data using L\'evy stable
source distributions. In this proceedings paper we omit the discussion of final-state
interactions, in particupar the effect of the Coulomb interaction. The handling of
this is described in detail in Ref.~\cite{Adare:2017vig}.

%We have carefully tested that this source model is in
%agreement with our data in all the transverse momentum regions reported here:
%ll the L\'evy fits were statistically acceptable, as indicated also later. 
%We note that using the method of L\'evy expansion of the
%correlation functions~\cite{Novak:2016cyc}, we have found that within errors all the
%terms that measure deviations from the L\'evy shape are consistent with zero.
%Hence we restrict the presentation of our results to the analysis of the
%correlation functions in terms of L\'evy stable source distributions.

\begin{figure}
\begin{center}
\includegraphics[width=0.6\textwidth]{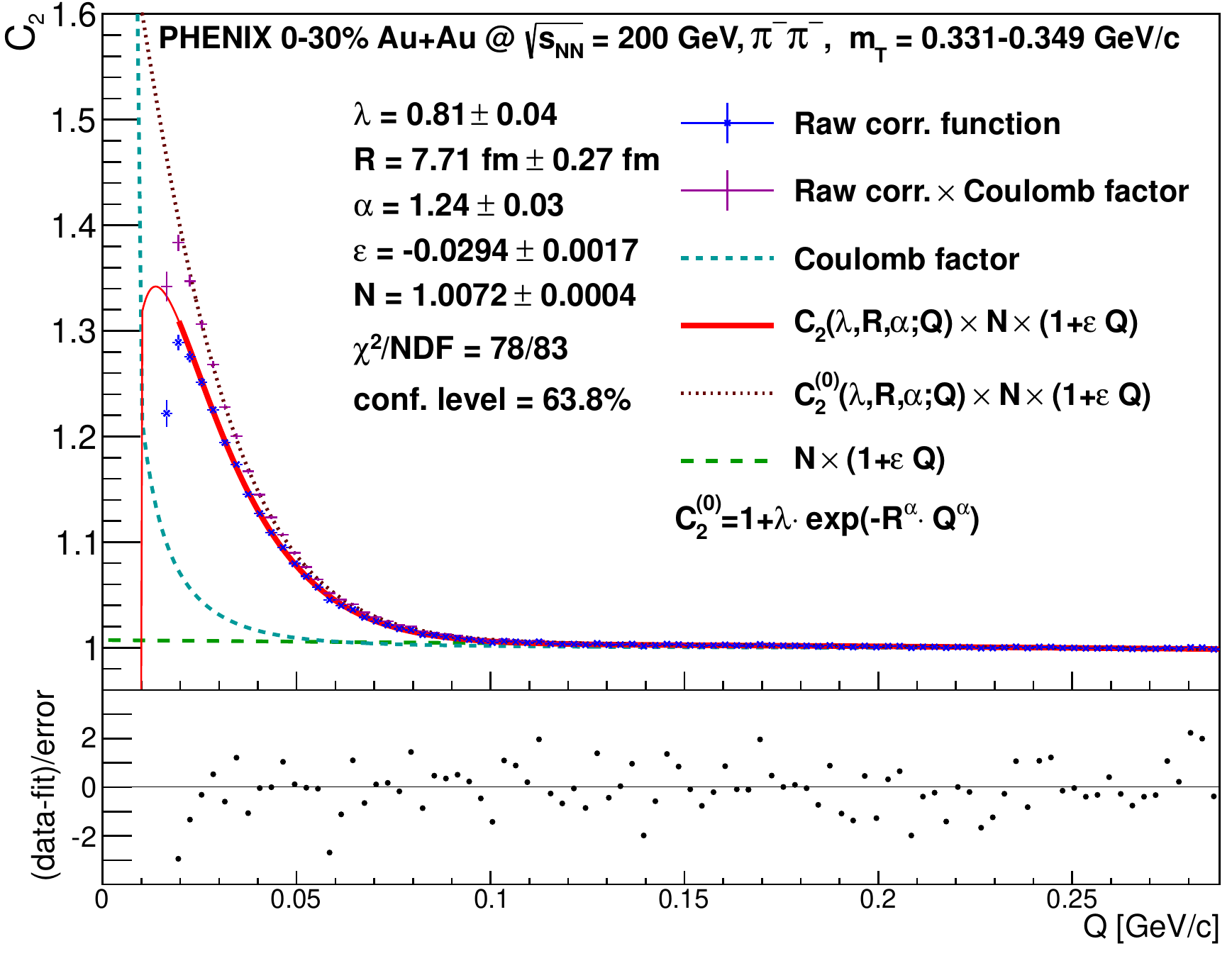}\vspace{-15pt}
\end{center}
\caption{Example fit of to a $\pi^-\pi^-$ correlation function, for $\mT = 0.331-0.349$ GeV$/c^2$.
The fit shows the measured correlation function and the complete
fit function, while a ``Coulomb-corrected" fit function $C^{(0)}(Q)$ is also shown, with the data multiplied
by $C^{(0)}/C^{\rm Coul}$.} \label{f:fit}
\end{figure}

\section{Results}

\begin{figure}
\begin{center}
\includegraphics[width=0.495\textwidth]{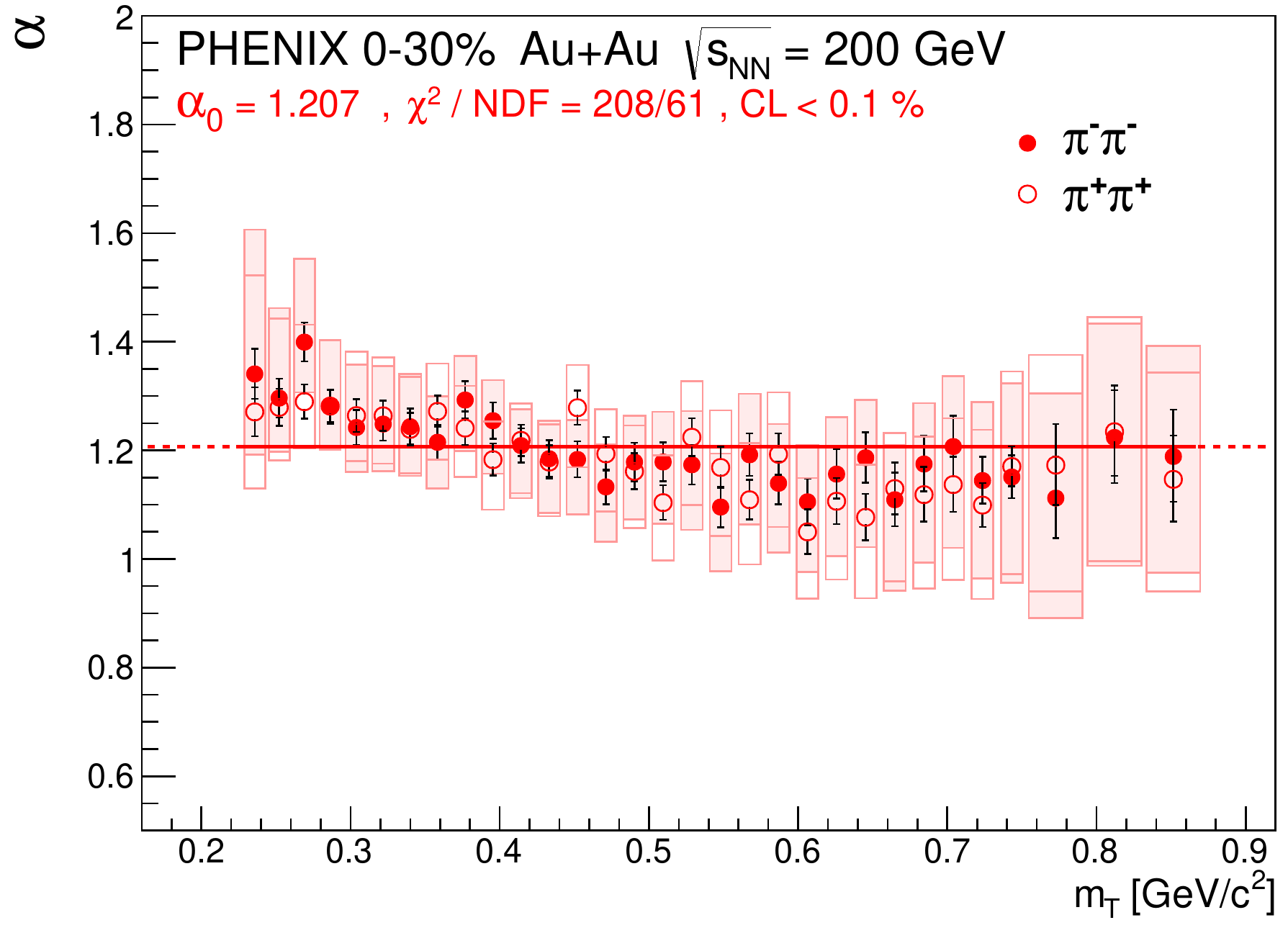}
\includegraphics[width=0.495\textwidth]{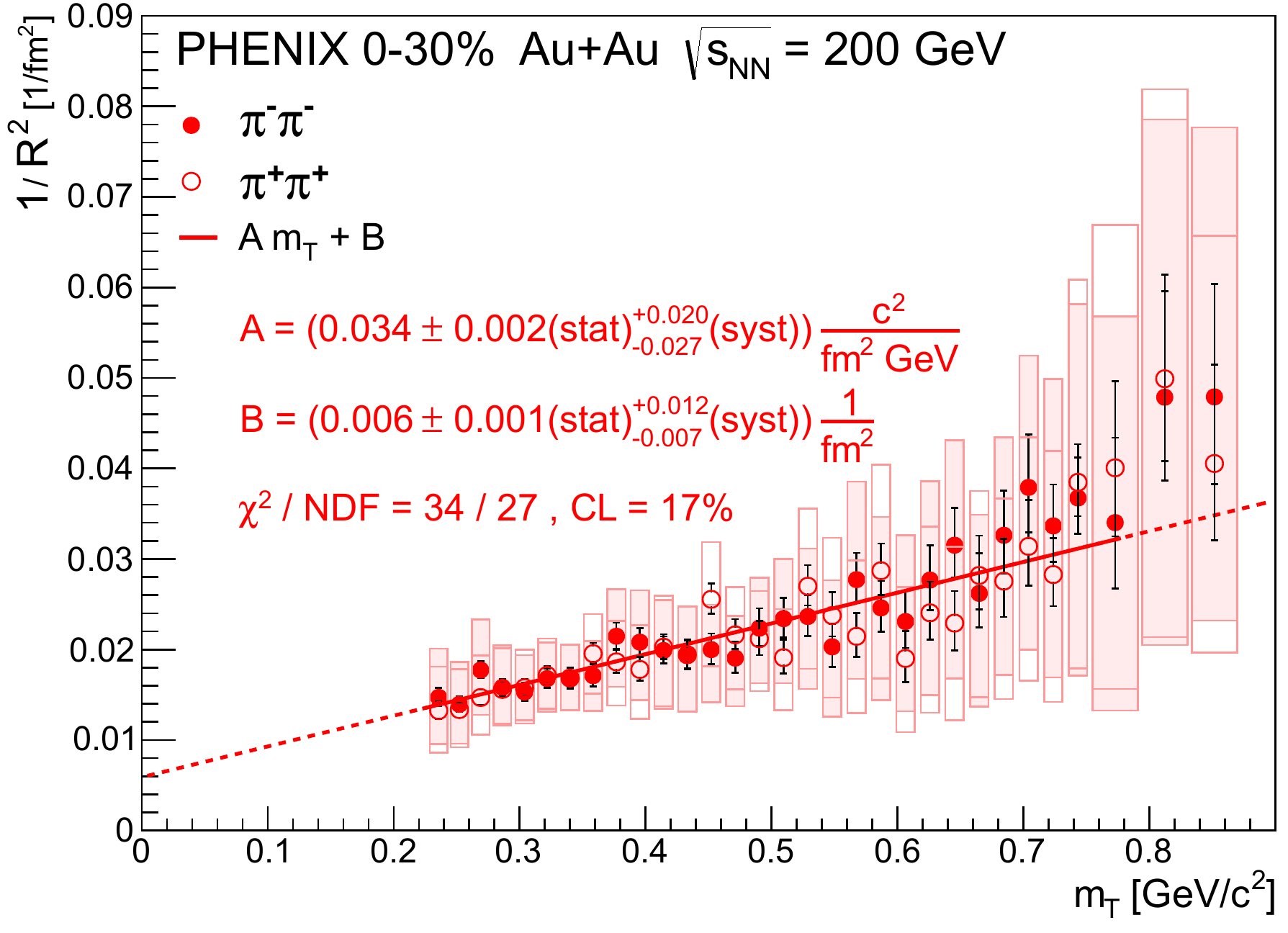}\\
\includegraphics[width=0.495\textwidth]{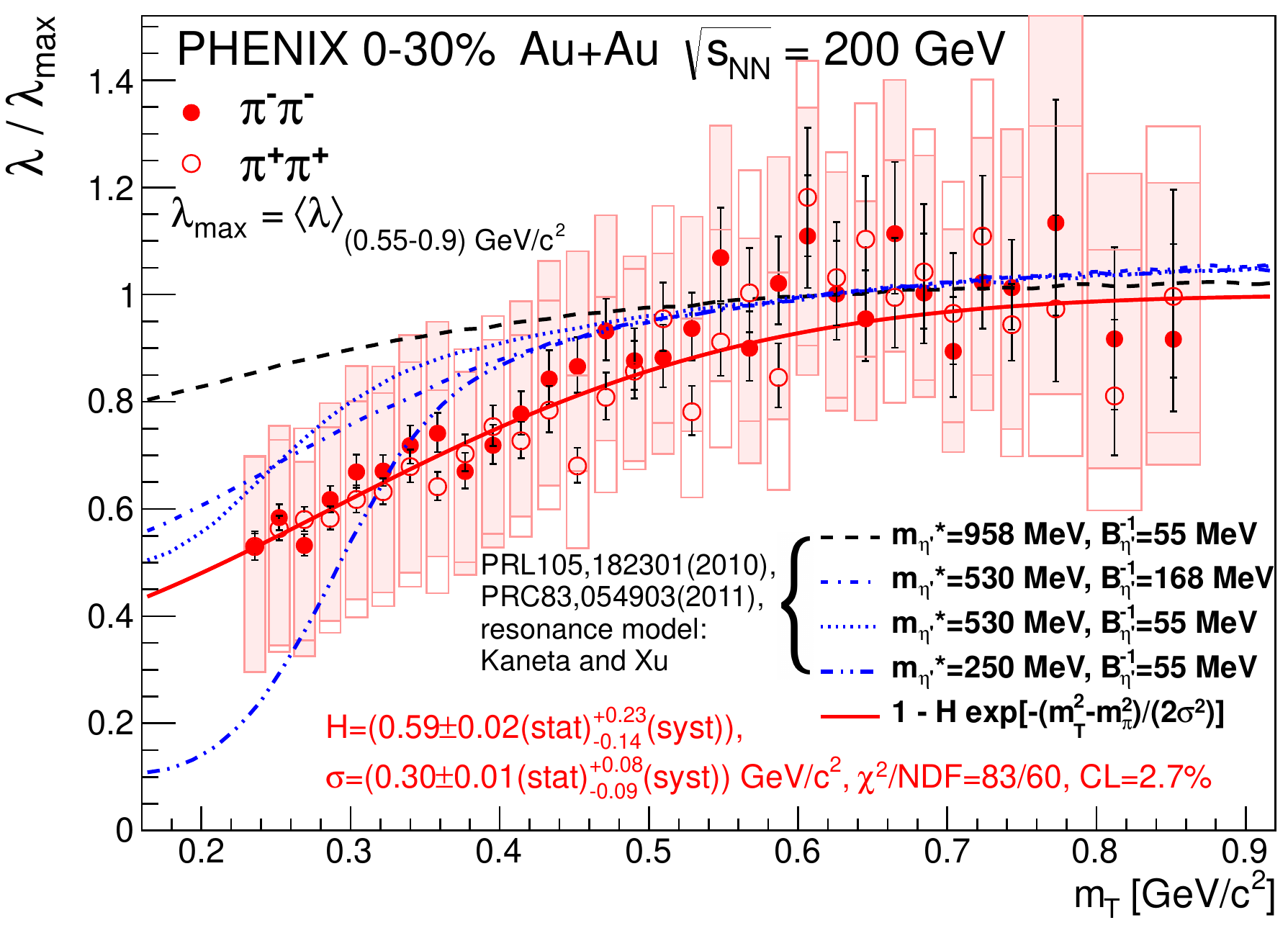}
\includegraphics[width=0.495\textwidth]{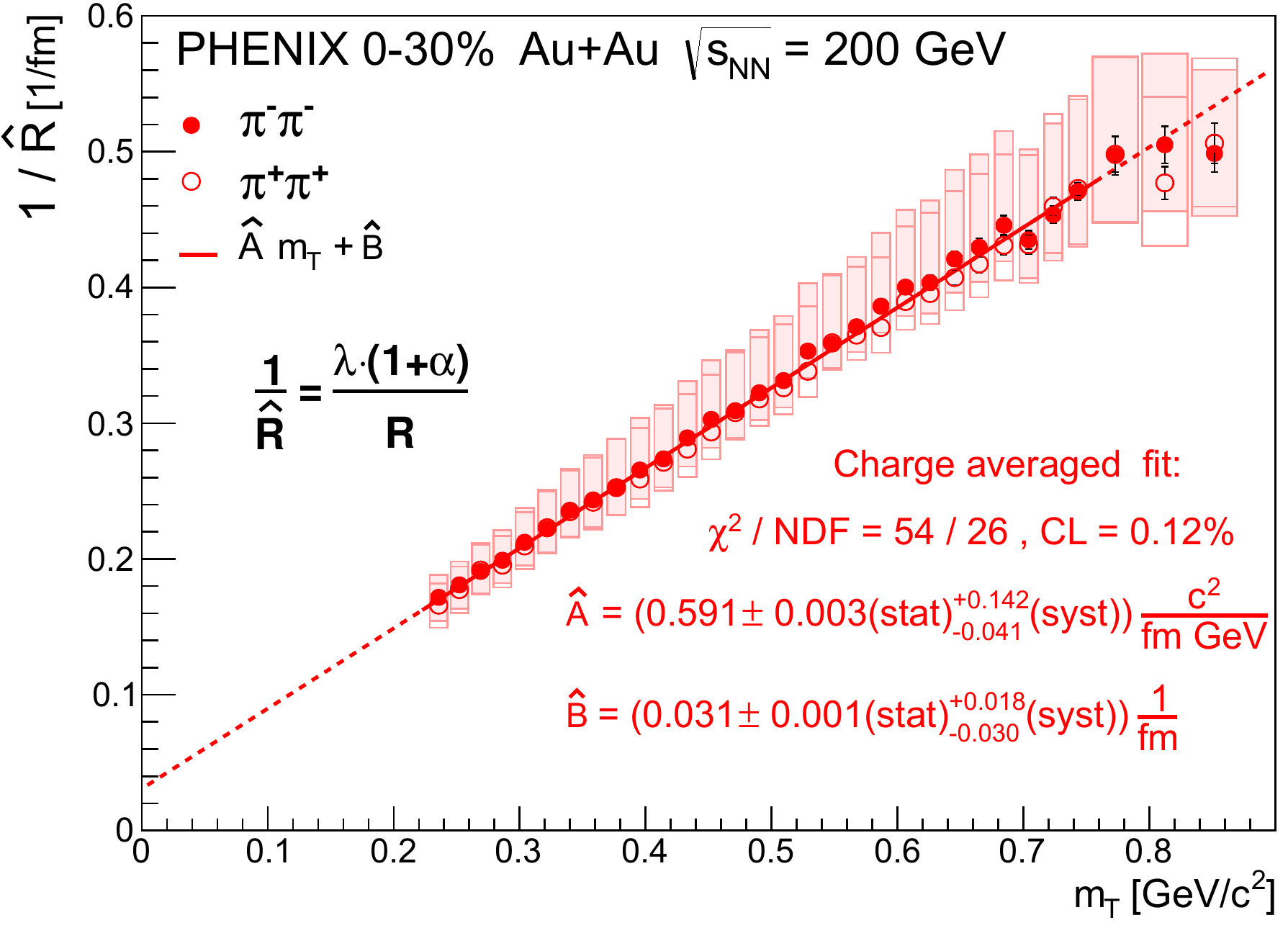}\vspace{-15pt}
\end{center}
\caption{Fit parameters versus average $m_T$ of the pair with statistical and symmetric systematic uncertainties shown as bars and boxes, respectively.}
\label{f:results}
\end{figure}

We analyzed \sqsntwo \auau collisions from the 2010 running 
period of the PHENIX experiment, selecting about 2.2 billion $0-30\%$ centrality events
from the recorded 7.3 billion Minimum Bias events. Note that in the original
conference presentation, the Minimum Bias data were presented (shown also e.g. in Ref.~\cite{Kincses:2016jsr}).
In this paper we present the final data from Ref.~\cite{Adare:2017vig}, which yield the same conclusion.
Two-particle correlation functions of $\pi^-\pi^-$ and $\pi^+\pi^+$ pairs (versus the momentum difference length in the longitudinally comoving
system, $Q$) were measured 31 $\mT$ bins ranging from 228 to 871 MeV$/c^2$ (where $\mT$
denotes the transverse momentum variable related to the average momentum $K$). We fitted
these correlation functions with the Coulomb-effect incorporated, based on Lévy-shaped sources,
as described in the previous section and in Ref.~\cite{Adare:2017vig}. Additionally, we introduced a linear background, as indicated
in Fig.~\ref{f:fit}, where an example fit is shown. The fits in all
$\mT$ bins and for both charges yield statistically acceptable descriptions of the measured
correlation functions, indicating that the fit parameters of $R$, $\alpha$ and $\lambda$ indeed
represent the measurements.

The $\mT$ dependence of the fit parameters is shown in Fig.~\ref{f:results}. We may observe that $\alpha$ is approximately
constant (within systematic uncertainties), and takes an average value of 1.207, being far from the Gaussian
assumption of $\alpha=2$, but also far from the conjectured $\alpha=0.5$ value at the critical point. The results are furthermore
incompatible with the exponential assumption of $\alpha=1$. We also
see, that despite being far from the hydrodynamic limit of Gaussian distributions, the hydro prediction of 
$1/R^2\simeq a+b\mT$ still holds. The correlation function strength $\lambda$ is shown after a normalization
by $\lambda_{\rm max}=\langle \lambda \rangle_{\mT=0.5-0.7 {\rm GeV}/c^2}$. This clearly indicates
a decrease at small $\mT$, which may be explained by resonance effects, and is in particular not 
incompatible with predictions based on a reduced $\eta'$ mass. We also show, that a new, empirically found scaling parameter
$\widehat{R}= R/(\lambda(1+\alpha))$ may be defined with decreased statistical uncertainties,
exhibiting a clear linear scaling with $\mT$.

{\it Acknowledgement: }
M. Cs. was supported by the New National Excellence program of the Hungarian Ministry of Human Capacities,
the NKFIH grant FK-123842 and the János Bolyai Research Scholarship.

%\bibliographystyle{prlstyl}
%\bibliography{../../../Master}

\end{document}